\documentclass[amsmath,amssymb,prl,final,showpacs,twocolumn]{revtex4}
\usepackage{bm,hyphenat,xspace}
\usepackage{graphicx,epsfig}

\newcommand{\Eq}{Eq.}

\newcommand{\Fig}{Fig.}
\newcommand{\Figs}{Figs.}
\newcommand{\Ref}{Ref.}
\newcommand{\Refs}{Refs.}

\newcommand {\mbf}[1]{{\mathbf{#1}}}

\newcommand {\mcu}{\mathcal{U}}

\newcommand {\mct}{\mathcal{T}}

\newcommand{\cm}{\mathrm{c\!\:\!.m\!\:\!.}}
\newcommand{\He}{{}^3\mathrm{He}}
\newcommand{\Hh}{{}^3\mathrm{H}}
\newcommand{\nH}{n\text{-}{}^3\mathrm{H}}
\newcommand{\pHe}{p\text{-}{}^3\mathrm{He}}
\newcommand{\pH}{p\text{-}{}^3\mathrm{H}}
\newcommand{\nHe}{n\text{-}{}^3\mathrm{He}}
\newcommand{\pd}{p\text{-}d}
\newcommand{\nd}{n\text{-}d}
\newcommand{\dd}{d\text{-}d}

\begin{document}

\title {\textit{Ab initio} four-body calculation of 
$n$-${}^3\mathrm{He}$, $p$-${}^3\mathrm{H}$, and $d$-$d$ scattering}
  
\author{A.~Deltuva} 
\email{deltuva@cii.fc.ul.pt}
\affiliation{Centro de F\'{\i}sica Nuclear da Universidade de Lisboa, 
P-1649-003 Lisboa, Portugal }

\author{A.~C.~Fonseca} 
\affiliation{Centro de F\'{\i}sica Nuclear da Universidade de Lisboa, 
P-1649-003 Lisboa, Portugal }

\received{ March 8, 2007}

\begin{abstract}
Four-body equations in momentum space are solved for neutron-$\He$,
proton-$\Hh$, and deuteron-deuteron scattering; all three reactions are 
coupled. The Coulomb interaction
between the protons is included using the screening and renormalization
approach as it was recently done for proton-deuteron and proton-$\He$
scattering.  Realistic interactions are used between nucleon pairs.
For the first time fully converged results for the observables pertaining 
to the six different elastic and transfer reactions are obtained
and compared with experimental data.   

\end{abstract}

 \maketitle

%%%%%%%%%%%%%%%%%%%%%%%%%%%%%%%%%%%%%%%%%%%%%%%%%%%%%%%%%%%%%%%%%%%%%%%%%%%%%%%

Modern studies of nuclear structure and reactions require \textit{ab
initio} calculations of the quantum many-body system using realistic 
interactions.
While at present there is a number of successful
\textit{ab initio} structure calculations using Green's function Monte Carlo
(GFMC)  methods  \cite{pieper:01a,pieper:02a} and No Core Shell Model
\cite{caurier:02a},
\textit{ab initio} studies of nuclear reactions are still limited to
few specific few-nucleon reactions. The difficulties associated with the
treatment of the long range Coulomb interaction restricted, until recently, 
precise calculations to neutron-deuteron $(\nd)$  elastic scattering and 
breakup \cite{gloeckle:96a}, and proton-deuteron $(\pd)$ \cite{kievsky:01a}, 
$\nH$ \cite{lazauskas:04a} and $\pHe$ \cite{viviani:01a} 
elastic scattering at low energy. The situation has now changed due
to the work in \Ref~\cite{deltuva:05a}. Using the ideas of
screening and renormalization~\cite{alt:80a} 
fully converged results for observables in $\pd$ elastic
scattering and breakup were obtained for a wide range of energies and
configurations using realistic force models. Other developments took
place recently using the  GFMC's approach~\cite{nollett:07a} to calculate the 
$n\text{-}{}^4\mathrm{He}$ phase shifts at energies well below the inelastic 
threshold.

In the present work we attempt to further extend \textit{ab initio}
calculations of the four-nucleon $(4N)$ system by solving the Alt, Grassberger 
and Sandhas (AGS) equations~\cite{grassberger:67} for the $\nHe$,
$\pH$  and $\dd$ reactions. This is the most complex scattering calculation 
that has been attempted so far since all three reactions are coupled and
involve both isospin $\mct = 0$ and $\mct = 1$ states, 
together with a very small admixture of $\mct = 2$ due to the
charge dependence of the hadronic and electromagnetic interactions. This work
is the continuation of two previous works for $\nH$~\cite{deltuva:07a}
and $\pHe$~\cite{fonseca:fb18} where the $4N$ scattering problem was
calculated  with the same level of accuracy as it already exists for 
three-nucleon $(3N)$ system. This means that calculations are carried out
without approximations on the two-nucleon $(2N)$  transition matrix 
(t-matrix) like in \Ref~\cite{fonseca:99a} or limitations on the choice 
of basis functions as in \Ref~\cite{fisher:06}. 
Therefore, after partial wave decomposition, the
AGS equations are three-variable integral equations that are solved
numerically without any approximations beyond the usual discretization
of continuum variables on a finite momentum mesh. Therefore
discrepancies with data may be attributed solely to the underlying $2N$
forces or lack of $3N$ forces.

The equations we solve  are based on the symmetrized four-body AGS
equations of \Ref~\cite{deltuva:07a}. In order to include the Coulomb
interaction we follow the methodology  of
\Refs~\cite{deltuva:05a,fonseca:fb18} and add to the nuclear $pp$
potential the screened Coulomb one $w_R$ that, in configuration space,
is given by
\begin{equation} \label{eq:wr}
w_R(r) = w(r) \, e^{-(r/R)^n},
\end{equation}
where  $w(r) = \alpha_e/r$ is the true Coulomb potential, 
$\alpha_e \simeq 1/137$ is the fine structure constant, 
and $n$ controls the smoothness of the screening; $n=4$ to 8 are the optimal 
values that ensure that $w_R(r)$ approximates well $w(r)$ for $r < R$
and simultaneously vanishes rapidly for $r>R$,
providing a comparatively fast convergence of the partial-wave expansion.
The screening radius $R$ must be considerably 
larger than the range of the strong interaction but from
the point of view of scattering theory $w_R$ is still of short range.
Therefore the equations  of Ref.~\cite{deltuva:07a} become $R$
dependent. The transition operators $\mcu^{\alpha\beta}_{(R)}$ where
$\alpha(\beta) = 1$ and 2 corresponds to initial/final  
$1+3$  and $2+2$ two-cluster states, respectively, satisfy the symmetrized
AGS equations
\begin{subequations}
\begin{align}  \nonumber
\mcu^{11}_{(R)}  = {}& -(G_0 \, t^{(R)}  G_0)^{-1}  P_{34} 
 - P_{34} \, U^1_{(R)}\, G_0 \, t^{(R)} G_0 \, \mcu^{11}_{(R)} \\
& + {U}^2_{(R)}   G_0 \, t^{(R)} G_0 \, \mcu^{21}_{(R)} , 
\label{eq:U11} \\  \nonumber
\mcu^{21}_{(R)}  = {}&  (G_0 \, t^{(R)}  G_0)^{-1} \, (1 - P_{34})
\\ & + (1 - P_{34}) U^1_{(R)}\, G_0 \, t^{(R)}  G_0 \,
\mcu^{11}_{(R)}  \label{eq:tildeU21}
\end{align}
\end{subequations}
for $1+3$ as the initial state and    
\begin{subequations}
\begin{align}  \nonumber
\mcu^{12}_{(R)}  = {}& (G_0 \, t^{(R)}  G_0)^{-1} 
 - P_{34} \, U^1_{(R)}\, G_0 \, t^{(R)} G_0 \, \mcu^{12}_{(R)} \\
& + {U}^2_{(R)}   G_0 \, t^{(R)} G_0 \, \mcu^{22}_{(R)} , 
\label{eq:U12} \\  \label{eq:tildeU22}
\mcu^{22}_{(R)} = {}& 
(1-P_{34}) U^1_{(R)}\, G_0 \, t^{(R)} G_0 \,\mcu^{12}_{(R)} 
\end{align}
\end{subequations}
for $2+2$ as the initial state. In both sets of equations $G_0$ is the four 
free particle Green's function and $t^{(R)}$ the 
$2N$ t-matrix derived from nuclear potential plus screened 
Coulomb between $pp$ pairs. The  operators $U^\alpha_{(R)}$ obtained from    
\begin{subequations} 
\begin{align}
\label{eq:U}
U^{\alpha}_{(R)} = {} & P_\alpha G_0^{-1} + 
P_\alpha \, t^{(R)}\, G_0 \, U^{\alpha}_{(R)}, \\
\label{eq:P}
P_1 = {} & P_{12}\, P_{23} + P_{13}\, P_{23}, \\
\label{eq:tildeP}  
P_2 = {} & P_{13}\, P_{24}, 
\end{align}
\end{subequations}
are the symmetrized AGS operators for the $1+(3)$ and $(2)+(2)$ subsystems
and $P_{ij}$ is the permutation operator of particles $i$ and $j$.
Defining the initial/final $1+(3)$ and $(2)+(2)$ states with relative
two-body momentum $\mbf{p}$
\begin{gather} \label{eq:phi1}
  | \phi_\alpha^{(R)} (\mbf{p}) \rangle = 
G_0 \, t^{(R)}  P_\alpha | \phi_\alpha^{(R)} (\mbf{p}) \rangle ,
\end{gather}
the amplitudes for all possible two-cluster transitions are obtained as
$\langle \mbf{p}_f | T^{\alpha\beta}_{(R)} | \mbf{p}_i \rangle = 
S_{\alpha\beta} \langle \phi_\alpha^{(R)} (\mbf{p}_f) | 
\mcu^{\alpha\beta}_{(R)} |\phi_\beta^{(R)} (\mbf{p}_i)\rangle $ with
$S_{11} = 3$, $S_{21} = \sqrt{3}$, $S_{22} = 2$ and $S_{12} = 2\sqrt{3}$. 

In close analogy with $\pd$ elastic scattering, the full scattering amplitude,
when calculated between initial and final $\pH$ or $\dd$ states,
 may  be decomposed as follows
\begin{equation}\label{eq:U11R}
T^{\alpha \beta}_{(R)} = t_{\alpha R}^{\cm} \; \delta_{\alpha \beta}
+ [T^{\alpha \beta}_{(R)} - t_{\alpha R}^{\cm} \; \delta_{\alpha \beta}],
\end{equation}
with the long-range part $t_{\alpha R}^{\cm}$ being the two-body
t-matrix derived  from the screened Coulomb potential of the form
\eqref{eq:wr} between the  proton and the center of mass (c.m.)  of 
$\Hh$ for $\alpha= 1$ and between the c.m. of both deuterons for $\alpha = 2$.
The remaining is the Coulomb-distorted  short-range part 
$[T^{\alpha \beta}_{(R)} - t_{\alpha R}^{\cm} \; \delta_{\alpha \beta}]$ 
as demonstrated in \Refs~\cite{alt:80a,deltuva:prep}. Applying the
renormalization procedure, i.e., multiplying both sides of
\Eq~(\ref{eq:U11R}) by the renormalization factors
$[Z_R^{\alpha}]^{-\frac{1}{2}}$ on the left and
$[Z_R^{\beta}]^{-\frac{1}{2}}$ on the right~\cite{deltuva:05a,alt:80a},
in the $R \to \infty$ limit, yields the full transition amplitude in the
presence of Coulomb
\begin{gather}      \label{eq:T} 
  \begin{split}
    \langle \mbf{p}_f| T^{\alpha \beta} |\mbf{p}_i \rangle  = {} &
    \langle \mbf{p}_f| t_{\alpha C}^{\cm} |\mbf{p}_i \rangle \;
\delta_{\alpha \beta}  %\\    & 
+ \lim_{R \to \infty} \left\{ [Z_R^{\alpha}]^{- \frac12}
\right.  \\ 
    & \times  \left.\langle \mbf{p}_f| [ T^{\alpha \beta}_{(R)} -
 t_{\alpha R}^{\cm} \; \delta_{\alpha \beta} ] |\mbf{p}_i
\rangle [Z_R^{\beta}]^{- \frac12} \right\},
    \end{split}
\end{gather}  
where the $[Z_R^{\alpha}]^{-1} \langle \mbf{p}_f| 
t_{\alpha R}^{\cm} |\mbf{p}_i \rangle $ 
converges (in general, as a distribution) to the exact Coulomb amplitude 
$\langle \mbf{p}_f| t_{\alpha C}^{\cm} |\mbf{p}_i \rangle$ 
between the proton and the c.m. of the $\Hh$ nucleus (between the c.m. of both 
deuterons) and therefore is replaced by it. 
The renormalization factor is employed in the partial-wave dependent form 
as in \Refs~\cite{deltuva:05a,fonseca:fb18}
\begin{gather} 
\label{eq:ZR}
Z_R^{\alpha} = e^{- 2i ( \sigma_L^{\alpha} - \eta_{LR}^{\alpha}) }
\end{gather} 
with the diverging screened Coulomb $\pH \; (\dd)$ phase shift
$\eta_{LR}^{\alpha}$ corresponding to standard boundary conditions
and the proper Coulomb one $\sigma_L^{\alpha}$ referring to the
logarithmically distorted proper Coulomb boundary conditions.
Obviously, there is no long-range Coulomb force in the $\nHe$ states;
in that case $\langle \mbf{p}_f| t_{\alpha C}^{\cm} |\mbf{p}_i \rangle = 
\langle \mbf{p}_f| t_{\alpha R}^{\cm} |\mbf{p}_i \rangle = 0$,
and $\sigma_L^{\alpha} = \eta_{LR}^{\alpha} = 0$.
The second term in \Eq~\eqref{eq:T},  after renormalization by 
$[Z_R^{\alpha}]^{- \frac{1}{2}} \; [Z_R^{\beta}]^{-\frac{1}{2}}$, 
represents the Coulomb-modified
nuclear short-range amplitude. It has to be calculated numerically,
but, due to its short-range nature, the $R \to \infty$ 
limit is reached with sufficient accuracy at finite screening
radii $R$ as demonstrated in \Refs~\cite{deltuva:05a,fonseca:fb18}
for $\pd$ and $\pHe$ scattering. As found there,  one needs larger 
values of $R$  at lower energies, making the convergence of the results more
difficult to reach. Nevertheless, for $E_p > 1$ MeV or $E_d > 1$ MeV the
method leads to very precise results. Depending on the reaction and the
energy, we obtain fully converged results for $\nHe$,
$\pH$,  and $\dd$ observables with $R$ ranging from 10 to 20 fm.

\begin{figure}[!]
\begin{center}
\includegraphics[scale=0.55]{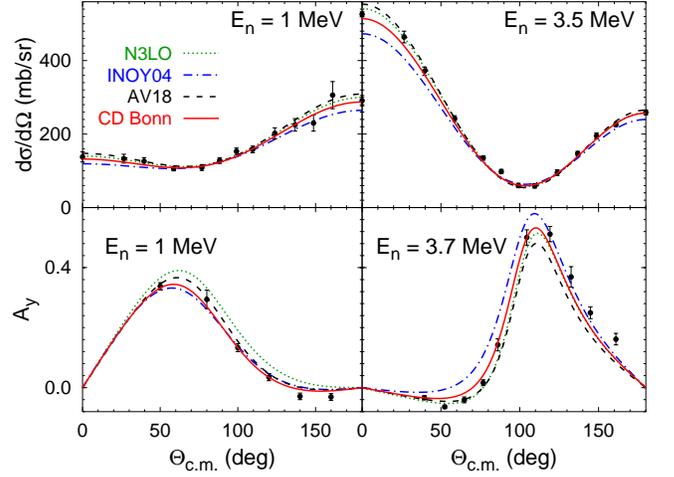}
\end{center}
\caption{\label{fig:nh} (Color online)
Differential cross section and neutron analyzing power of elastic
$\nHe$ scattering at 1, 3.5, and 3.7 MeV neutron lab energy. 
Results obtained with potentials
CD Bonn (solid curves), AV18 (dashed curves),
INOY04 (dashed-dotted curves), and N3LO (dotted curves) are compared.
The cross section data  are from \Ref~\cite{seagrave:60},
$A_y$ data are from \Ref~\cite{jany:88} at 1 MeV and from 
\Ref~\cite{klages:85} at 3.7 MeV.}
\end{figure}

\begin{figure}[!]
\begin{center}
\includegraphics[scale=0.48]{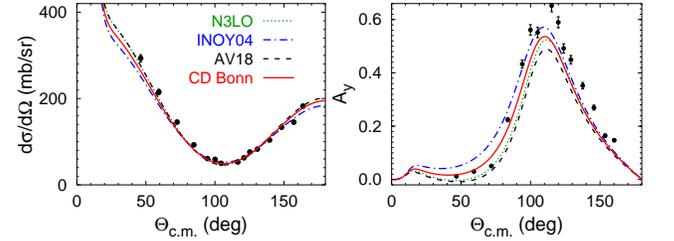}
\end{center}
\caption{\label{fig:pt} (Color online)
Differential cross section and proton analyzing power of elastic
$\pH$ scattering at 4.15 MeV proton lab energy. 
Curves as in \Fig~\ref{fig:nh}.
The data  are from \Ref~\cite{kankowsky:76}.}
\end{figure}

\begin{figure}[!]
\begin{center}
\includegraphics[scale=0.48]{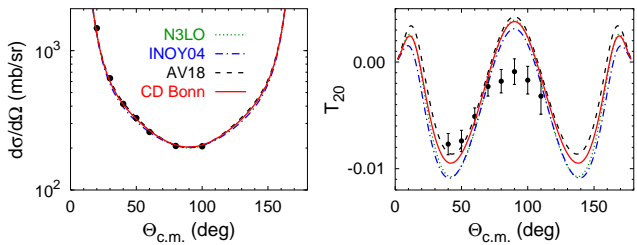}
\end{center}
\caption{\label{fig:dd} (Color online)
Differential cross section and deuteron tensor analyzing power $T_{20}$
of elastic $\dd$ scattering at 3 MeV deuteron lab energy. 
Curves as in \Fig~\ref{fig:nh}.
The cross section data  are from \Ref~\cite{blair:48a} and $T_{20}$ data
are from \Ref~\cite{crowe:00a}.}
\end{figure}

\begin{figure}[!]
\begin{center}
\includegraphics[scale=0.55]{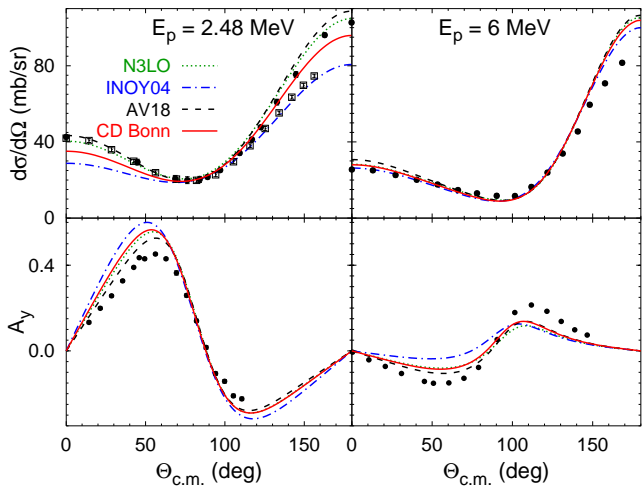}
\end{center}
\caption{\label{fig:pt-nh} (Color online)
Differential cross section and proton analyzing power of 
$p+\Hh \to n + \He$ reaction at 2.48 and 6 MeV proton lab energy. 
Curves as in \Fig~\ref{fig:nh}.
The cross section data  are from \Refs~\cite{drosg:78} (circles) and
\cite{jarvis:56} (squares) at 2.48 MeV, and  from \Ref~\cite{wilson:61}
at 6 MeV. $A_y$ data are from \Ref~\cite{doyle:81} at 2.48 MeV
and from \Ref~\cite{jarmer:74} at 6 MeV.}
\end{figure}

\begin{figure}[!]
\begin{center}
\includegraphics[scale=0.55]{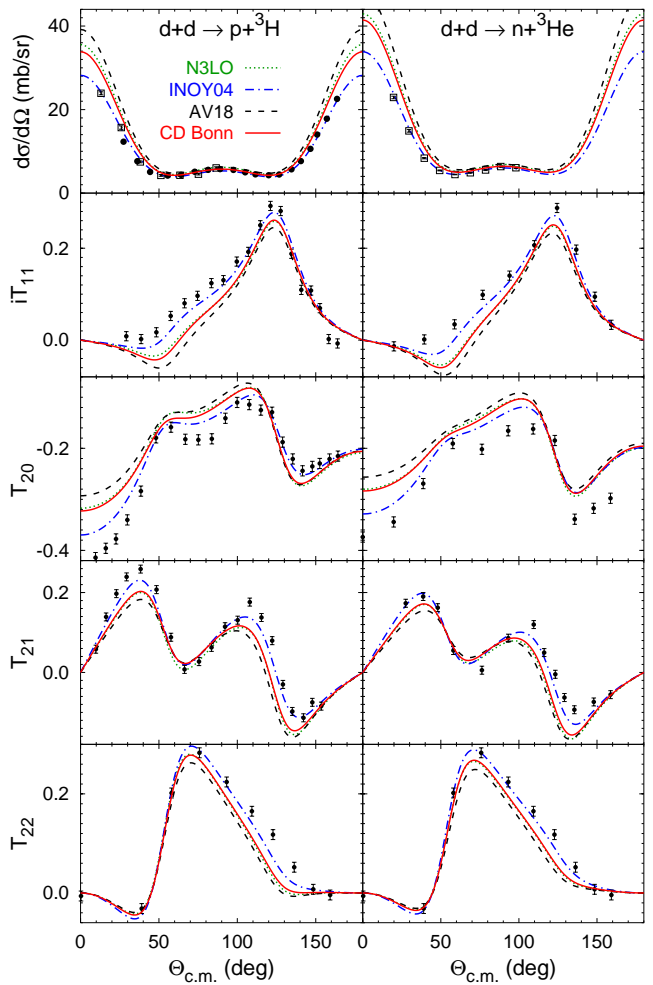}
\end{center}
\caption{\label{fig:dd-ptnh} (Color online)
Differential cross section and deuteron analyzing powers of 
$d+d \to p+\Hh$ and  $d+d \to n+\He$ reactions at 3 MeV deuteron lab energy. 
Curves as in \Fig~\ref{fig:nh}.
The cross section data  are from \Refs~\cite{blair:48b} (squares) 
and \cite{gruebler:72a} (circles). Analyzing power data are from
\Ref~\cite{gruebler:72a} for $d+d \to p+\Hh$ and from
\Ref~\cite{dries:79a} for $d+d \to n+\He$.}
\end{figure}

The results are also fully converged with respect to the partial-wave
expansion. The calculations include isospin-singlet $2N$ partial waves
with  total angular momentum $I \leq 4$ and  isospin-triplet $2N$
partial waves  with orbital angular momentum $l_x \leq 6$, $3N$ partial
waves with spectator orbital angular momentum $l_y \leq 6$ and 
total angular momentum $J \leq \frac{13}{2}$, $4N$ partial waves
with 1+3 and 2+2 orbital angular momentum $l_z \leq 6$, resulting
up to about 15000 channels for fixed $4N$ total angular momentum and parity.
Initial/final two-cluster states with orbital angular
momentum $L \leq 5$ are included for the calculation of observables.
For some reactions, e.g., $\nHe$ elastic scattering, the partial-wave
convergence is considerably faster, allowing for a reduction in the
employed angular momentum cutoffs.

The two-nucleon interactions we use are charge-dependent 
(CD) Bonn~\cite{machleidt:01a}, AV18~\cite{wiringa:95a}, 
inside nonlocal outside Yukawa (INOY04) potential by
Doleschall~\cite{doleschall:04a}, and the one derived from chiral
perturbation theory at next-to-next-to-next-to-leading order
(N3LO)~\cite{entem:03a}.
In~\Figs~\ref{fig:nh} through ~\ref{fig:dd-ptnh} we show the results
for six different reactions.  The most remarkable findings are: a) the
excellent agreement with the data for the calculated $\dd$ elastic
differential cross section; b) the lack of a large $A_y$ deficiency in
$\nHe$ and $\pH$ unlike what was observed before in
$\pHe$~\cite{fonseca:fb18,fisher:06}; c) the overall description of
$d+d \to n+\He$ and $d+d \to p+\Hh$ data, given its complex structure.

Of all the potentials we use only INOY04 represents both $\nHe$ and
$\pH$ thresholds in the correct position (-7.72 MeV and -8.48 MeV,
respectively). For this reason it is perhaps not surprising to see that
this potential gives rise to the best description of $d+d \to
n+\He$ and $d+d \to p+\Hh$ data, including the differential cross
section. Nevertheless, even if the correct position of thresholds is an
important factor, it is certainly not the only one that matters since in
$\nHe$ elastic scattering and $p+\Hh \to n+\He$ INOY04 gives rise to 
results that are poorer than those obtained with other potentials. 

The largest disagreements with data are observed in a very small $\dd$ 
elastic $T_{20}$ at $90^\circ$, and in the $p+\Hh \to n+\He$ 
proton analyzing power and differential cross section at forward and
backward angles at low energies. At $E_p = 2.48$ MeV we find two sets
of data that are inconsistent at backward angles but where this work
does not help to resolve given the wide range of results obtained with
the potentials we use. At $E_p = 6$ MeV the results are almost
independent of the choice of potentials but they all miss the
data at backward angles. The problems associated with $\nHe$ and $\pH$
reactions may be associated with the $0^+$ excited state of the alpha
particle that sits between the two thresholds and whose position
changes with the potential we choose. 

\begin{figure}[!]
\begin{center}
\includegraphics[scale=0.48]{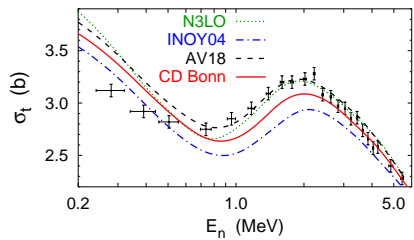}
\end{center}
\caption{\label{fig:nhtot} (Color online)
$\nHe$ total cross section as function of the neutron lab energy.
Curves as in \Fig~\ref{fig:nh}.
The data  are from \Ref~\cite{nhetot}.}
\end{figure}

Finally in \Fig~\ref{fig:nhtot} we show the total $\nHe$ cross section
as a function of energy. Again we find an agreement with data that, 
depending on the potential, is either similar (INOY04) or better 
(N3LO and AV18) than observed in $\nH$ \cite{deltuva:07a}.
As in $\nH$ total cross section \cite{deltuva:07a}, the curves pertaining to 
AV18 and N3LO calculations exchange position at about the same excitation 
energy relative to the  $\nH$  threshold (E $\simeq$ 1.3 MeV) but now N3LO 
leads to the highest total cross section at low energy instead of at the 
resonance peak. The reason for this behavior is unclear at this time, 
but may reside in the relative position of the $0^-$ $(^3{\rm P}_0)$ 
inelastic resonance~\cite{fonseca:02} that is associated with the $0^-$ 
$\mct = 0$  state at about 0.4 MeV excitation above the $\nHe$ threshold. 
 
In conclusion, we show for the first time results of  \textit{ab initio} 
calculations involving four-nucleon reactions initiated by either  
$\nHe$,  $\pH$, and $\dd$. Realistic  nuclear potentials 
are used between hadron pairs plus the Coulomb interaction between protons. 
Although specific observables show discrepancies with data that need 
further investigation, namely the inclusion of three-nucleon forces, 
the results  give a reasonable overall description of the data for all 
six reactions that show an improvement relative to what is found for 
$\mct = 1$ observables \cite{deltuva:07a,fonseca:fb18}. Work is in progress 
to study the effect of three-nucleon forces through the inclusion of 
$\Delta$ degrees of freedom.
 
%\clearpage

A.D. is supported by the Funda\c{c}\~{a}o para a Ci\^{e}ncia e a
Tecnologia (FCT) grant SFRH/BPD/14801/2003
and A.C.F. in part by the FCT grant POCTI/ISFL/2/275.

%%%%%%%%%%%%%%%%%%%%%%%%%%%%%%%%%%%%%%%%%%%%%%%%%%%%%%%%%%%%%%%%%%%%%%%%%%%%%

%\clearpage
%%%%%%%%%%%%%%%%%%%%%%%%%%%%%%%%%%%%%%%%%%%%%%%%%%%%%%%%%%%%%%%%%%%%%%%%%%%%%
%\bibliographystyle{prsty3}
%\bibliography{abbrev,pre80,80-89,90-99,200x,clmb,ad,4N,hann,book,numerics}

\end{document}